\documentclass[a4paper,11pt,unpublished,allowtoday]{quantumarticle}
\pdfoutput=1
\usepackage[utf8]{inputenc}
\usepackage[english]{babel}
\usepackage[T1]{fontenc}
\usepackage{amsmath}
\usepackage{amssymb}
\usepackage{mathtools, nccmath}
\usepackage{setspace}
\usepackage{hyperref}
\usepackage{outlines}
\usepackage{fancyvrb}
\usepackage{physics}
\usepackage{comment}
\usepackage{bbold}
\usepackage{cancel}
\usepackage{graphicx}
\usepackage[caption=false]{subfig}
\usepackage[usenames, dvipsnames]{color}
\usepackage{soul}
\usepackage{tensor}
\usepackage{amsthm}

\usepackage{caption}

\newcommand{\floor}[1]{\left\lfloor #1 \right\rfloor}

\usepackage[numbers,sort&compress]{natbib}  
\usepackage{tikz}
\usepackage{quantikz}
\usepackage{lipsum}
\usepackage{tkz-graph}
\usetikzlibrary{hobby,backgrounds,calc,trees}
\usepackage{xcolor}
\usetikzlibrary{positioning}
\usepackage{ulem}
\usepackage{yquant}
\usepackage{caption}
\useyquantlanguage{groups}
\usepackage{makecell}
\usepackage{tikz,esvect,tikz-3dplot}
\usetikzlibrary{3d,calc,intersections}

\usetikzlibrary{shapes.geometric, arrows}

\tikzstyle{arrow} = [thick,->,>=stealth]

\usepackage{algorithm,algpseudocode}
\algnewcommand{\To}{\textbf{To }}
\algnewcommand\Input{\item[\textbf{Input:}]}%
\algnewcommand\Output{\item[\textbf{Output:}]}%

\begin{document}

\title{An iterative transversal CNOT decoder}

\date{\today}

\author{Kwok Ho Wan}
\orcid{0000-0002-1762-1001}
\email{khw1496((avocado))gmail((donut))com}
\affiliation{Universal Quantum Ltd, Gemini House, Mill Green Business Estate, Haywards Heath, RH16 1XQ, United Kingdom}
\affiliation{Blackett Laboratory, Imperial College London, South Kensington, London SW7 2AZ, United Kingdom}
\thanks{Current position \& affiliation: Academic Visitor at Blackett Laboratory, Imperial College London, SW7 2AZ, UK}
\author{Mark Webber}
\orcid{0000-0002-8674-9439}
\affiliation{Universal Quantum Ltd, Gemini House, Mill Green Business Estate, Haywards Heath, RH16 1XQ, United Kingdom}

\author{Austin G. Fowler}
\affiliation{Google Inc., Santa Barbara, 93117 CA, USA}

\author{Winfried K. Hensinger}
\orcid{0000-0001-8329-438X}
\affiliation{Sussex Centre for Quantum Technologies, University of Sussex, Brighton, BN1 9RH, United Kingdom}
\affiliation{Universal Quantum Ltd, Gemini House, Mill Green Business Estate, Haywards Heath, RH16 1XQ, United Kingdom}

\begin{abstract}
Modern platforms for potential qubit candidates, such as trapped ions or neutral atoms, allow long range connectivity between distant physical qubits through shuttling. This opens up an avenue for transversal logical CNOT gates between distant logical qubits, whereby physical CNOT gates are performed between each corresponding physical qubit on the control and target logical qubits. However, the transversal CNOT can propagate errors from one logical qubit to another, leading to correlated errors between logical qubits. We have developed a multi-pass iterative decoder that decodes each logical qubit separately to deal with this correlated error. We show that under circuit-level noise and only $\mathcal{O}(1)$ code cycles, a threshold can still persist, and the logical error rate will not be significantly degraded, matching the sub-threshold logical error rate scaling of $p^{\floor{\frac{d}{2}}}$ for a distance $d$ rotated surface code. 
\end{abstract}

\maketitle

\section{Introduction}
Quantum error correction (QEC) is essential for realising fault-tolerant quantum computation. The threshold theorem states that if the error rate per quantum gate is below a certain threshold, arbitrarily long quantum computations become possible \cite{gottesman2009introductionquantumerrorcorrection}. The surface code, known for its high error threshold and ability to correct both bit-flip and phase-flip errors through local interactions, is a leading QEC candidate. It arranges qubits on a two-dimensional lattice and uses stabilizer measurements for error detection. The minimum weight perfect matching (MWPM) algorithm is an effective decoding method for the surface code \cite{pymatchingv2,higgott2023sparseblossomcorrectingmillion,wu2023fusionblossomfastmwpm}.

\begin{figure}[!h]
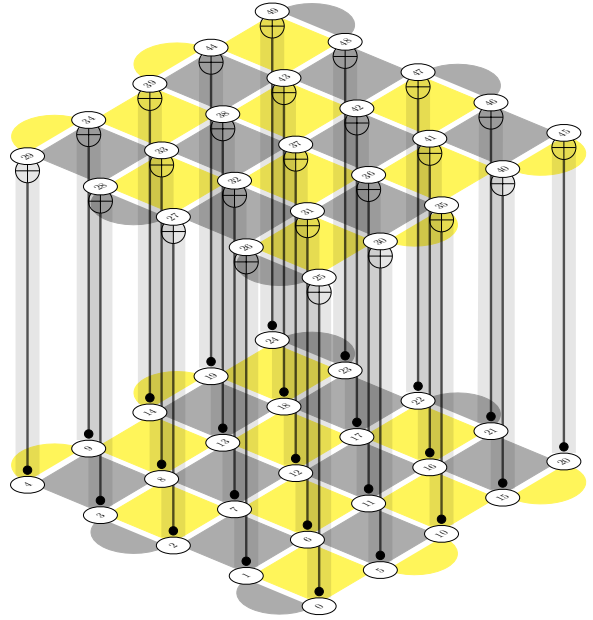

    \centering
        \resizebox{0.95\linewidth}{!}{

} 
    \caption{Two patches/copies of distance $5$ rotated surface codes. A transversal logical CNOT operation is performed by applying physical CNOT gates between the corresponding physical qubits of the two patches.}
    \label{fig:two_patches}
\end{figure}

Quantum computing hardware varies in connectivity capabilities. Superconducting qubits typically have fixed, limited connectivity which is imposed by the chip topology, while platforms like trapped ions, neutral atoms, and photonics allow for long-range connectivity. For example, in the trapped ion paradigm, ion shuttling operations enable arbitrary connectivity \cite{webber2020efficientqubitroutingglobally}. The variability of connectivity influences QEC implementation and efficiency across different systems. 

Lattice surgery is a common technique for performing logical operations in the surface code while relying only on nearest neighbour operations. It involves merging and splitting logical qubits to facilitate operations such as the logical CNOT gate. The spacetime complexity of lattice surgery scales with the code distance, $\mathcal{O}(d^3)$, where $d$ is the code distance. One common example lattice surgery implementation of logical CNOT uses $2d$ rounds of error detection and three logical qubits, one of which is an ancilla patch for mediating the merging and splitting operations \cite{Horsman_2012,chatterjee2024latticesurgerydummies}. Note that all Clifford operations can be done in $\mathcal{O}(1)$ time given sufficient qubits \cite{fowler2013timeoptimalquantumcomputation}.

The transversal CNOT gate offers an alternative approach. It requires performing a physical CNOT operation between every corresponding physical qubit of the logical qubits. For $N$ data qubits per logical qubit, $N$ CNOT operations are needed, often involving long-range interactions. The transversal CNOT deterministically maps Pauli errors between patches: X errors are mapped from control to target, and Z errors are mapped from target to control. This mapping, if uncorrected, leads to spurious detection events (DEs) which will greatly degrade logical qubit performance if standard decoding methods like MWPM are used.

Recent research has focused on developing specialized decoders for transversal CNOT gates~\cite{PRXQuantum.2.020341,Bluvstein2024,cain2024correlated,zhou2024algorithmicfaulttolerancefast,hangleiter2024faulttolerantcompilingclassicallyhard,kim2024transversal,SahayYT2024,hetényi2024creating}. Our work significantly expands upon a concept briefly mentioned\footnote{As a long footnote.} in~\cite{PRXQuantum.2.020341}, by developing a multi-pass iterative decoder that addresses the correlated errors spread by the transversal CNOT. This decoder processes each logical patch separately in multiple passes to correct these correlated errors. Previously, correlated decoding methods have been demonstrated to exponentially suppress errors for transversal CNOT circuits~\cite{cain2024correlated}, relying on detectors conditioned on measurement outcomes from multiple logical qubits. That work compares a scalable Belief-HUF decoder with a non-scalable MLE decoder \cite{cain2024correlated}. In contrast, our method uniquely enables MWPM to decode each logical qubit independently, with additional logic interleaved to suppress correlated error propagation. The resulting logical error rates compare favorably with those from memory-equivalent circuits. As both runtime and achieved logical error rates can vary significantly across decoding approaches, we argue that exploring structurally distinct decoder architectures is essential. Our method maintains compatibility with scalable MWPM frameworks while explicitly addressing correlated error propagation—capabilities not simultaneously realized in existing methods.

Modern qubit platforms supporting long-range connectivity make transversal CNOT gates feasible, potentially improving resource estimates for quantum computing. Our multi-pass iterative decoder demonstrates that, even under circuit-level noise and only $\mathcal{O}(1)$ code cycles, a threshold can persist without significant degradation of the logical error rate. This matches the sub-threshold logical error rate scaling of $p^{\lfloor d/2 \rfloor}$ for a distance $d$ rotated surface code.

This paper presents an analysis of our multi-pass iterative decoder for transversal CNOT gates, including its design, implementation, and performance characteristics. 

Specifically, we illustrate a simple method to separately decode two different patches of surface codes that had undergone transversal CNOT operations, we assume:
\begin{enumerate}
    \item error-free and instantaneous transversal CNOT operations and
    \item standard circuit-level depolarising noise model on all other gates - SD6 \cite{ErrorCorrectionZoo}.
\end{enumerate}

We argue that the error-free transversal operation should not significantly impact the overall behaviour of the logical error rates, as the realistic transversal CNOT error is expected to contribute only a marginally higher circuit-level noise error rate \cite{pymatchingv2} for some hardware platforms \cite{webber2020efficientqubitroutingglobally}. Furthermore, the idealized transversal CNOT operation suffices for the primary purposes of our investigation here, i.e. the performance of a new approach to decoding. A detailed analysis into hardware realistic connectivity error models and its impact on the performance of the transversal CNOT operation relative to lattice surgery is in preparation \cite{tvslscnot}.

Our work has significant implications for resource estimation in quantum advantage applications, where there exists a trade-off between run times and the number of physical qubits. Previous work \cite{fowler2019lowoverheadquantumcomputation,Gidney_2019,tan2024satscalpellatticesurgery} has assumed lattice surgery operations and demonstrated the benefits of fast code cycle times when other factors are held constant. For instance, trapped ion hardware, with code cycle times potentially three orders of magnitude slower than superconducting qubits, would traditionally require substantially more physical qubits to achieve comparable quantum advantage run times \cite{Webber_2022}. This comparison assumes the same QEC methods and physical error rates across both platforms.

However, transversal operations offer a substantial advantage, bringing the scaling of the spacetime overhead of computation down from $\mathcal{O}(d^3)$ to $\mathcal{O}(d^2)$. By leveraging this advantage along with enhanced connectivity and higher baseline fidelities on physical gates, slower hardware architectures may be able to achieve competitive run-times without excessive physical qubit overhead. Our findings suggest that the integration of transversal CNOT gates and specialized decoding techniques could significantly impact the resource landscape across diverse quantum computing platforms, potentially altering the balance between speed and qubit count in the pursuit of quantum advantage.

\section{The repetition code}
In the following sections, we illustrate our iterative transversal CNOT decoding method with the repetition code. Due to the surface code's symmetry in suppressing Pauli-X and -Z error, it is sufficient to study the repetition code for the intuition behind our method. For simplicity, we also assume no mid-circuit errors for the visualisation of this decoding procedure. However, it should be noted that this method easily generalises to accommodate mid-circuit errors (see appendix \ref{appendix:surface_code_implementation_and_pauli_frame_ingerence}).

\begin{figure}[h]
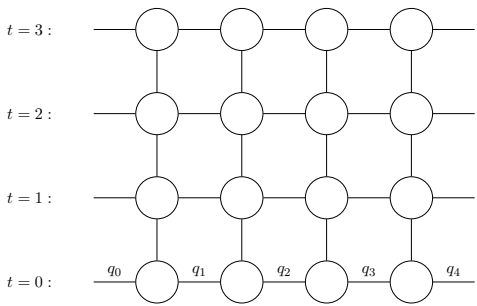

\centering
    \resizebox{0.85\linewidth}{!}{

    }    
    \caption{\label{fig:rep_code_1} Distance $5$ repetition code (can tolerate $\floor{d/2}=2$ errors before failing) matching graph for 4 time steps, assuming no mid-circuit errors (no diagonal edges in the matching graph). }
\end{figure}

A distance $d$ repetition code has code words $\ket{\bar{0}}=\ket{0}^{\otimes d},\ket{\bar{1}}=\ket{1}^{\otimes d}$ and can be represented with the following stabiliser generators, $\mathcal{G}$, and logical operators $\bar{X}$ and $\bar{Z}$:
\begin{equation}
    \begin{split}
        \mathcal{G}=\Bigg\langle \bigcup_{j=0}^{d-2} Z_jZ_{j+1} \Bigg\rangle, \ \bar{X} = \prod_{i=0}^{d-1} X_i \ , \ \bar{Z} = Z_k \ ,
    \end{split}
\end{equation}
where $Z_j$($X_j$) are Pauli-Z(X) operator acting on physical qubit $j$. If we were to repeatedly measure the stabiliser generators of this code, we can detect and correct for Pauli-X errors only. In the particular example in figure~\ref{fig:rep_code_1}, we initialise the repetition code and measure the parity checks for 3 rounds before subsequently measuring all the data qubits. We can then construct the syndrome of this quantum memory experiment \cite{fowler2019lowoverheadquantumcomputation} (in this case a classical code). For $t=1,2$, the syndrome/DEs can be constructed by the calculating the modulo 2 difference ($\oplus$) of the same parity check measurement separate by one step in time. 

Figure \ref{fig:rep_code_1} is a matching graph for this quantum memory experiment with 3 rounds of stabiliser measurement followed by data qubit measurement. The horizontal edges at each time slice of this graph represent the possibility of bit-flip errors on the qubits and vertical edges between time slices correspond to the possibility of measurement errors. Detectors are the nodes of this graph, and will be coloured if the parity of the measurements associated with the detector differs from that expected (a DE). 

\section{Correlated errors}
A CNOT operation inherently spreads errors between the control and target qubits. X errors will flow from the control to target qubit (shown in equation \ref{eq:X_spread}) and Z errors will flow from the target to control qubit (see equation \ref{eq:Z_spread}).

\begin{equation}
\label{eq:X_spread}
\begin{split}
\begin{tikzpicture}
\begin{yquantgroup}
\registers{
qubit {} q[2];
}
\circuit{
X q[0];
cnot q[1] | q[0];
}
\equals
\circuit{
cnot q[1] | q[0];
X -;
}
\end{yquantgroup}
\end{tikzpicture}
\\
\begin{tikzpicture}
\begin{yquantgroup}
\registers{
qubit {} q[2];
}
\circuit{
X q[1];
cnot q[1] | q[0];
}
\equals
\circuit{
cnot q[1] | q[0];
X q[1];
}
\end{yquantgroup}
\end{tikzpicture}
\end{split}
\end{equation}

\begin{equation}
\label{eq:Z_spread}
\begin{split}
\begin{tikzpicture}
\begin{yquantgroup}
\registers{
qubit {} q[2];
}
\circuit{
Z q[0];
cnot q[1] | q[0];
}
\equals
\circuit{
cnot q[1] | q[0];
Z q[0];
}
\end{yquantgroup}
\end{tikzpicture}
\\
\begin{tikzpicture}
\begin{yquantgroup}
\registers{
qubit {} q[2];
}
\circuit{
Z q[1];
cnot q[1] | q[0];
}
\equals
\circuit{
cnot q[1] | q[0];
Z -;
}
\end{yquantgroup}
\end{tikzpicture}
\end{split}
\end{equation}

For simplicity, in this discussion we focus on the X errors only and note that by symmetry, the argument is valid for the Z errors with the control and target modes. The transversal CNOT will deterministically map X errors from the control to target patch. If these propagated errors are not corrected for, they will cause spurious DEs appearing on the target right after the transversal CNOT.  Ultimately, this has the overall effect of  a degraded logical error rate. However, if we can first decode the control patch, and have a suitable guess for the locations of the X errors, we can use that to cancel out the propagated X errors on the target patch. The motivation behind our decoder is to use a conventional memory experiment decoder for a single logical qubit memory experiment, which decodes each logical qubit separately, updates the syndrome data and Pauli-frame of each individual logical qubit and then propagate that information before decoding again iteratively. This process repeats until the Pauli-frames converge to a stable state. The number of iterations required is a function of the circuit structure, where simple circuits only ever require one iteration. 

\section{Iterative decoder for the repetition code}
We shall illustrate our approach with two examples concerning the repetition code. The first example (Example 1) involves a single transversal CNOT, requiring only a single-pass decoder, whereas the multiple CNOT version (Example 2) requires multiple rounds of iterative decoding. 

We use the following notation and conventions to separate different detectors and edges by colour.
\begin{itemize}
    \item ``Natural'' errors (edges) and detectors (nodes) are coloured {\color{red} red}, \resizebox{0.2\linewidth}{!}{

    }, they are the errored edges and detectors that occur genuinely, and are not due to the propagation effects of the CNOT.

    \item Propagated errors (edges) and detectors (nodes) are coloured {\color{brown} brown}, \resizebox{0.2\linewidth}{!}{
    %
    }, they are the errored edges and detectors that are caused solely by the propagation of errors due to the CNOT.

    \item Decoded matched edges are coloured {\color{blue} blue}, they are the edges returned by the MWPM decoding algorithm, 
    \resizebox{0.2\linewidth}{!}{
    %
    }, in this case the blue decoded edge had correctly corrected for the red corrupted edge.

    \item The stored propagated edges are coloured {\color{green} green}, they are the recorded edges \textit{perceived} to be propagated by the error spreading effects of the CNOT, 
    \resizebox{0.2\linewidth}{!}{
    %
    }, in this case the green propagated edge had correctly corrected for the brown propagated edge.
\end{itemize}

The core part of the method relies on propagating Pauli errors between the patches and toggling the presence of DEs. For example with X errors, the Pauli frame of the control patch directly before the CNOT operation is determined, and propagated in the round following the CNOT into a separate stored list corresponding to the target patch. These propagated errors are then used to toggle the presence of DEs in the target patch. The procedure acts to recover the detector pattern corresponding to only the natural errors. The logical observable is determined by combining the decoded Pauli frame with the propagated Pauli frame. 

Each CNOT is addressed in this manner sequentially. For more complex circuits, like those with alternating directions of CNOT operations, more than one iteration may be required, where each iteration involves addressing every CNOT in the circuit, first undoing the impact of this particular CNOT from the prior iteration, and then again performing the propagation of Pauli errors and detectors. The termination clause is when no DEs are toggled. For circuits with two alternating direction CNOT operations, at most two iterations are ever required. Further details illustrating the iterative method is found the appendix \ref{appendix:psuedocode_2_patch}.

\subsection{Example 1: Single pass decoder}
\begin{figure}[!h]
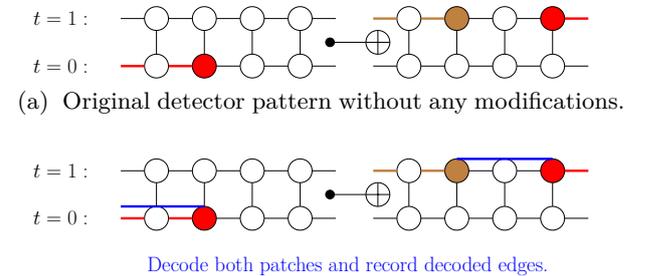

\centering
    \subfloat[\label{fig:ex1_og} Original detector pattern without any modifications.]{{
    \resizebox{0.95\linewidth}{!}{

    }
    }}
    \qquad
    \subfloat[\label{fig:ex1_step2} Decode both logical qubits and record the blue decoded edges, we can see that we have arrived at incorrectly decoded edges.]{{
    \resizebox{0.95\linewidth}{!}{
    %
    }
    }}
    \caption{\label{fig:example_1_a} Figures \ref{fig:ex1_og} and \ref{fig:ex1_step2} show the decoding procedure before the propagation of the Pauli-frame from the left to the right logical qubit in order to reduce the number of spurious DEs resulting from the transversal logical CNOT. We can observe that in this case, we have decoded and arrived at incorrect edges.}
\end{figure}

\begin{figure}[!h]
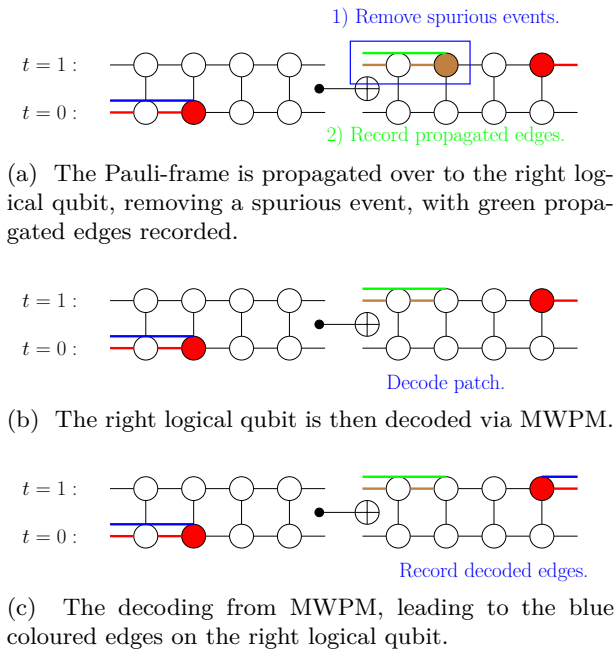

\centering
    \qquad
    \subfloat[\label{fig:ex1_step3} The Pauli-frame is propagated over to the right logical qubit, removing a spurious event, with green propagated edges recorded.]{{
    \resizebox{0.95\linewidth}{!}{

    }
    }}
    \qquad
    \subfloat[\label{fig:ex1_step4} The right logical qubit is then decoded via MWPM.]{{
    \resizebox{0.95\linewidth}{!}{
    %
    }
    }}
    \qquad
    \subfloat[\label{fig:ex1_step5} The decoding from MWPM, leading to the blue coloured edges on the right logical qubit.]{{
    \resizebox{0.95\linewidth}{!}{
    %
    }
    }}
\caption{\label{fig:example_1_b} Steps in figure \ref{fig:ex1_step3} to figure \ref{fig:ex1_step5} show the decoding procedure required to decode and propagate the Pauli-frame from the left logical qubit to the right logical qubit.}
\end{figure}

In figures \ref{fig:example_1_a} and \ref{fig:example_1_b}, chronologically in time from figure \ref{fig:ex1_og} to figure \ref{fig:ex1_step5}, we show a method to modify DEs and record propagated edges to decode a single logical transversal CNOT acting between two logical qubits (left and right).

Firstly, in figure \ref{fig:ex1_og}, we see a detector pattern, and have annotated the propagated detectors and edges with the brown nodes and edges respectively, and also labelled the errored qubits with red edges for illustrative purposes. However in an actual experiment, this information is not available.

In figure \ref{fig:ex1_step2}, we decode the left and right logical qubits and record the blue decoded edges in figure \ref{fig:ex1_step2}. In this case, decoding has chosen some edges incorrectly.

In order to decode correctly, we reset the blue decoded edges on the right patch, then, the predicted Pauli-frame for the left logical qubit is propagated to the right logical qubit in figure \ref{fig:ex1_step4}, removing a spurious DE and recording the propagated edges in green  at $t=1$. We then proceed to decode the right logical qubit with MWPM and record its decoded edges in figure \ref{fig:ex1_step5}. 

With this syndrome modification and Pauli-frame propagation procedure, we are able to correctly identify the ``natural errors'' and the propagated errors with all the spurious DEs removed. For an instantaneous and error-free transversal CNOT operation, the logical error rate of failure after the application of a transversal CNOT should be as close as possible to the memory experiment case. This can be interpreted as a memory experiment obfuscated with a transversal CNOT.

\subsection{Example 2: Multi-pass decoder}
We will now show a more complicated example which includes two alternating transversal CNOTs. Multiple iterations of frame propagation will be required. In the case of two alternating transversal CNOT's, we observe that an X error spreading from control to target in the first instance will be reflected back after the second transversal CNOT. If not correctly decoded, this would lead to an amplification of errors that will severely affect the overall logical error rate. 

\subsubsection{$1^{\text{st}}$ iteration}
\begin{figure}[!h]
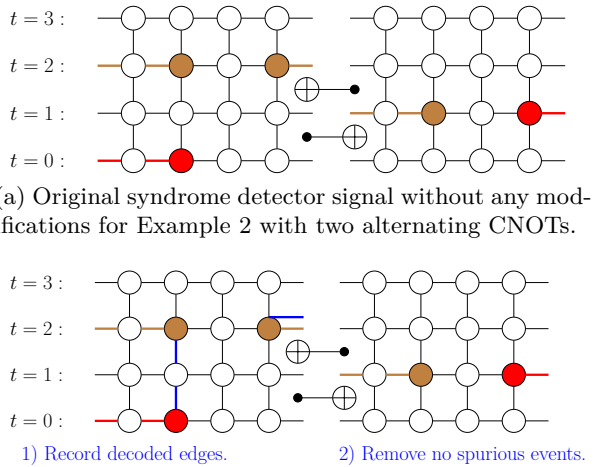

\centering
    \subfloat[\label{fig:ex2_og}Original syndrome detector signal without any modifications for Example 2 with two alternating CNOTs.]{{
    \resizebox{0.95\linewidth}{!}{

    }}}
    \qquad     
    \subfloat[\label{fig:ex2_step_2}We decode the left patch and propagate the Pauli-frame to the right patch, in this case, no errors were detected, so no spurious events or propagation of green edges were formed on the right patch.]{{
    \resizebox{0.95\linewidth}{!}{
    %
    }}}
\caption{\label{fig:example_2_part_a} The original syndrome detection pattern of the two alternating CNOT example, following by the first iteration's first Pauli-frame propagation through the first CNOT.}
\end{figure}

In the first iteration, we follow similar methods outlined in Example 1. The detector pattern from experiment can be seen in figure \ref{fig:ex2_og}, the series of two alternating transversal CNOTs with a round of syndrome extraction in between leads to the brown propagated edges and DEs. We first decode the left logical qubit and record its blue decoded edges in \ref{fig:ex2_step_2}. We then propagate the Pauli-frame at $t=0$ in the left patch to the right patch to remove spurious events, in this case no data (spatial) edges are matched in the left patch at $t=0$, hence no spurious events are propagated over to the right. These are the incorrectly decoded edges on the left, we shall see how iterating the process in the $2^{\text{nd}}$ iteration solves this.

\begin{figure}[!h]
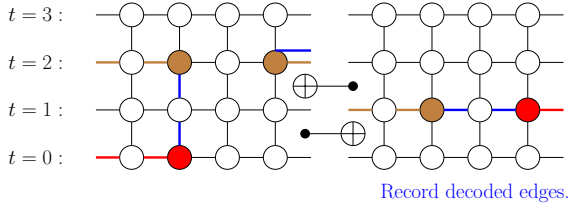
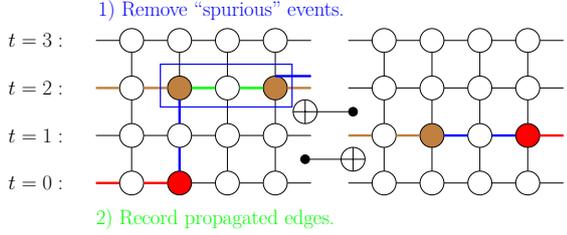

\centering
    \subfloat[\label{fig:ex2_step_4}Decode the logical qubit on the right and record its decoded blue coloured edges.]{{
    \resizebox{0.95\linewidth}{!}{
    %
    }}}
    \qquad     
    \subfloat[\label{fig:ex2_step_5} Using the Pauli-frame of the right patch up to $t=1$, propagated the frame over to the left, record the propagated edges and remove the spurious events. Note that the propagated edges and the removed spurious event are incorrect and will be fixed in the second iteration.]{{
    \resizebox{0.95\linewidth}{!}{
    %
    }}}
    
\caption{\label{fig:example_2_part_b} The procedure outlining the first iteration's second Pauli-frame propagation through the second CNOT.}
\end{figure}

\begin{figure}[!h]
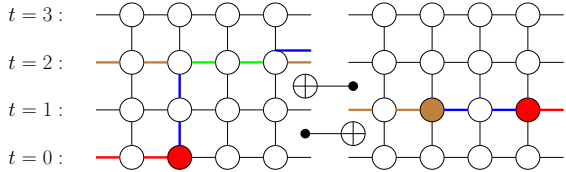

\centering
    \resizebox{0.95\linewidth}{!}{

    }
\caption{\label{fig:example_2_end_of_iter_1} The decoded and propagated edge record after the first iteration, note that we have removed two spurious DEs but at this stage introduced a logical error. This will be rectified in the second iteration.}
\end{figure}

Then, we decode the logical qubit on the right, leading to the blue decoded edges in figure \ref{fig:ex2_step_4}. Again these matched edges are incorrect, we will still propagate the frame over from the right at $t=1$ to the left at $t=2$, removing spurious events and recording the green decoded edges (figure \ref{fig:ex2_step_5}). At the end of the first iteration, we arrive at a series of matched edges as shown in figure \ref{fig:example_2_end_of_iter_1}. The pattern of DEs has changed, and this triggers another round of decoding.

\subsubsection{$2^{\text{nd}}$ iteration}
\begin{figure}[!h]
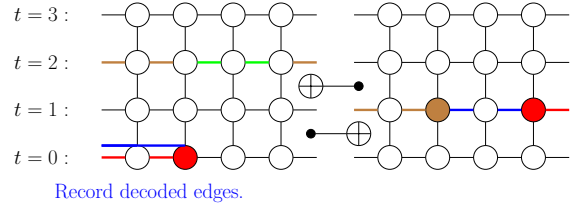
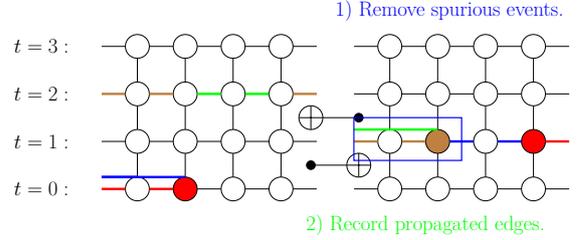

\centering    
    \subfloat[\label{fig:ex2_step_7}Reset the decoded blue edges on the left logical qubit and then re-decode.]{{
    \resizebox{0.95\linewidth}{!}{

    }}}
    \qquad     
    \subfloat[\label{fig:ex2_step_8}Propagate the Pauli-frame up till $t=0$ from the left logical qubit from figure \ref{fig:ex2_step_7} to the right logical qubit at $t=1$.]{{
    \resizebox{0.95\linewidth}{!}{
    %
    }}}
\caption{\label{fig:example_2_part_c} In the second iteration of the Pauli-frame through the first CNOT, we can see that we have correctly identified some more of the errored edges.}
\end{figure}

It's clear that a single iteration of frame propagation will not be sufficient to find good decoded edges in this case. In the second iteration, we reset the blue edges on the left qubit and decode again leading to new blue edges at $t=0$ (figure \ref{fig:ex2_step_7}). Propagate the Pauli-frame over to $t=1$ of the logical qubit on the right, removing the spurious events and record the newly propagated green edges (figure \ref{fig:ex2_step_8}). 

\begin{figure}[!h]
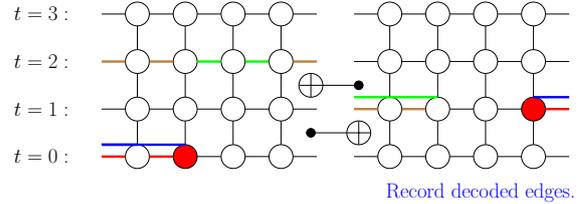
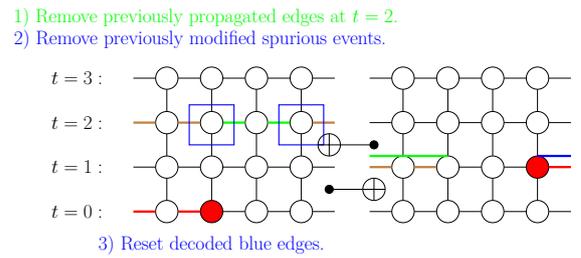

\centering   
    \subfloat[\label{fig:ex2_step_10} Reset the decoded blue edges on the right patch and re-decode.]{{
    \resizebox{0.95\linewidth}{!}{

    }}}
    \qquad    
    \subfloat[\label{fig:ex2_step_11} Firstly, we remove the previously modified spurious events and propagated green edge at $t=2$ on the left logical qubit from the previous iteration.]{{
    \resizebox{0.95\linewidth}{!}{
    %
    }}}
\caption{\label{fig:example_2_part_d} Crucial step in the second iteration of Pauli-frame propagation through the second CNOT, we need to undo detector modifications and propagated edges done in the first iteration.}
\end{figure}

\begin{figure}[!h]
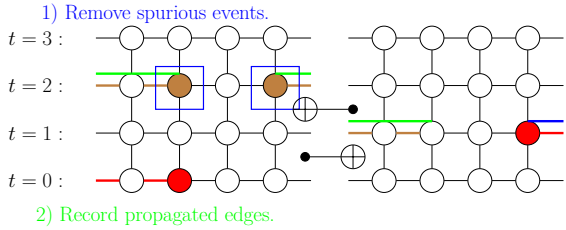
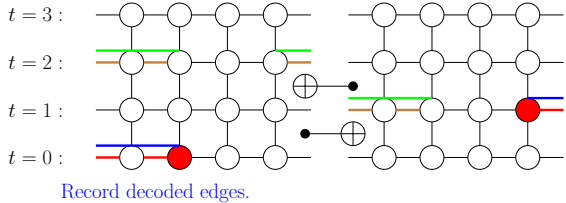

\centering       
    \subfloat[\label{fig:ex2_step_12} Re-propagate the green edges from the right patch to the left patch at $t=2$, removing new spurious events and record new propagated edges.]{{
    \resizebox{0.95\linewidth}{!}{
    %
    }}}
    \qquad     
    \subfloat[\label{fig:ex2_step_14} Finally, re-decode the left logical qubit and we have finally arrived at a pattern of detectors and set of Pauli frames that is stable warranting no further iteration.]{{
    \resizebox{0.95\linewidth}{!}{
    %
    }}}
\caption{\label{fig:example_2_part_e} Final frame propagation leading to all the correct corrupted edges being identified.}
\end{figure}

Next, the right logical qubit is decoded, leading to blue edges at $t=1$ in figure \ref{fig:ex2_step_10}. Before propagating the frame from $t=1$ on the right to left, we need to remove the previously propagated edges and spurious events at $t=2$ on the left from the first iteration (see figure \ref{fig:ex2_step_11}). We then reset the decoded blue edges on the left, before following the conventional frame propagation shown in figure \ref{fig:ex2_step_12} to \ref{fig:ex2_step_14}. 

After the second iteration, we have correctly identified all the propagated edges and errored edges. Please refer to the appendix \ref{appendix:psuedocode_2_patch} for an outline of the process.

\section{Surface code simulations}
The procedure illustrated in the previous sections can be easily generalised to patches of surface code after applications of transversal CNOT operations. We aim to study and simulate the minimally interesting cases. Firstly, we characterise a transversal CNOT operation memory experiment with one round of syndrome extraction before and after it. Then we study the next minimally interesting case with 2 alternating transversal CNOTs, with the iterative procedure and varying number of rounds of syndrome extraction between the CNOTs. All the logical error rates show close agreement with the memory experiment logical error rates.

We use Stim \cite{gidney2021stim} to construct the detector error model and sample from the syndrome extraction circuits undergoing the SD6 circuit-level noise model. We then subsequently use PyMatching to decode each logical qubit's syndromes separately with an equivalent single patch memory experiment matching graph.

We represent arbitrary number of rounds of syndrome extraction that is performed as follows:
\begin{equation}
    \label{eq:cir:SE_def}
    \begin{tikzpicture}[on/.style={red, very thick}]
    \begin{yquant}[on/.style={style=red, control style={very thick}}]
    qubit {$\ket{\bar{0}}$} q;
    box {$x$ SE} (q[0]);
    box {$\hat{U}$} (q[0]);
    box {$y$ SE} (q[0]);
    measure q;
    setstyle {} q[0];
    \end{yquant}
    \end{tikzpicture} \ \ ,
    \end{equation} 
\noindent the quantum circuit in equation \ref{eq:cir:SE_def} shows one logical qubit undergoing $x$ rounds of syndrome extraction(s), before a unitary operation $\hat{U}$, followed by another $y$ rounds of syndrome extraction(s) before reading it out in the computational basis.

\subsection{Single transversal CNOT}

\begin{figure}[!h]
\centering
\resizebox{0.5\linewidth}{!}{
    \begin{tikzpicture}
    \begin{yquant}
    qubit {} q;
    qubit {} q[+1];
    box {$1$ SE} (q[0]-q[1]);
    cnot q[1] | q[0];
    box {$1$ SE} (q[0]-q[1]);
    measure q;
    \end{yquant}
    \end{tikzpicture}
    }
\caption*{A single CNOT circuit with one round of syndrome extraction before and after, followed by measurement of the logical observables.}
\end{figure}
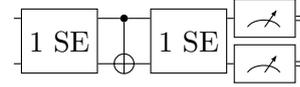

\begin{figure}[!h]
    \centering
    \includegraphics[width=1\linewidth]{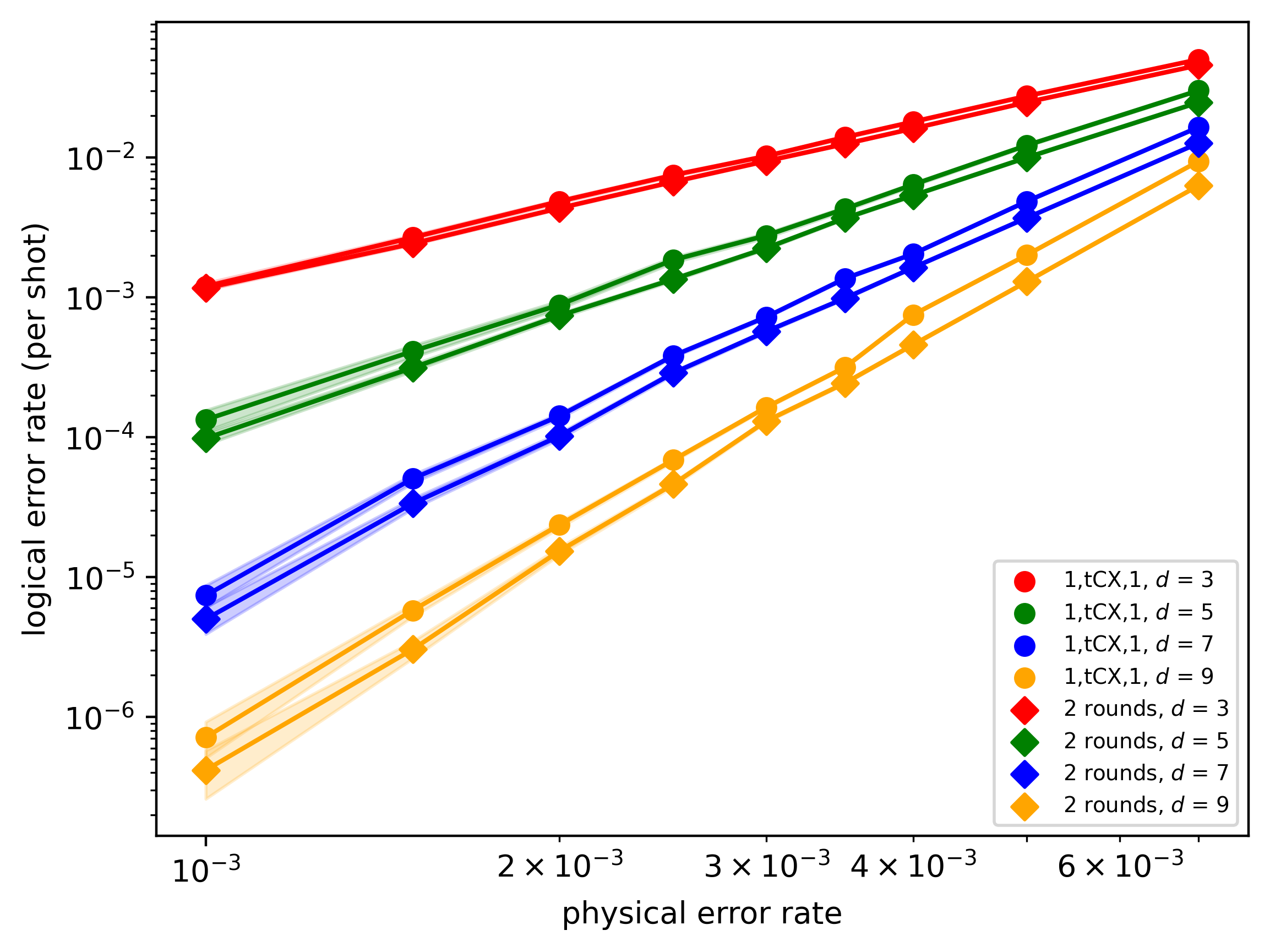}
    \caption{Surface code simulations with a single transversal CNOT sandwich between one round of syndrome extraction before and after before reading out all the data qubits. At every distance, the transversal CNOT logical error rate matches the memory experiment results closely.}
    \label{fig:1_tCX_1}
\end{figure}

In figure \ref{fig:1_tCX_1} we analyse the single CNOT case, where only one iteration is ever required. We show close agreement between the CNOT circuit with propagating decoder versus the memory equivalent, which corresponds to two logical qubits with no CNOT operations and the same number of total rounds, i.e. two rounds, decoded with standard independent MWPM. This represents a lower-bound on performance quality. The scaling relationship with physical error rate is preserved. 

\subsection{Two alternating transversal CNOTs}

\begin{figure}[!h]
\centering
\resizebox{0.7\linewidth}{!}{
    \begin{tikzpicture}
    \begin{yquant}
    qubit {} q;
    qubit {} q[+1];
    box {$1$ SE} (q[0]-q[1]);
    cnot q[1] | q[0];
    box {$n_r$ SE} (q[0]-q[1]);
    cnot q[0] | q[1];
    box {$1$ SE} (q[0]-q[1]);
    measure q;
    \end{yquant}
    \end{tikzpicture}
    }
\caption*{A two-alternating CNOT circuit with one round of syndrome extraction before, a variable number of rounds between, and one after, followed by measurement of the logical observables.}
\end{figure}
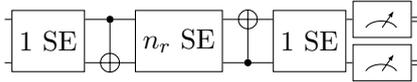

\begin{figure}[!h]
    \centering
    \includegraphics[width=1\linewidth]{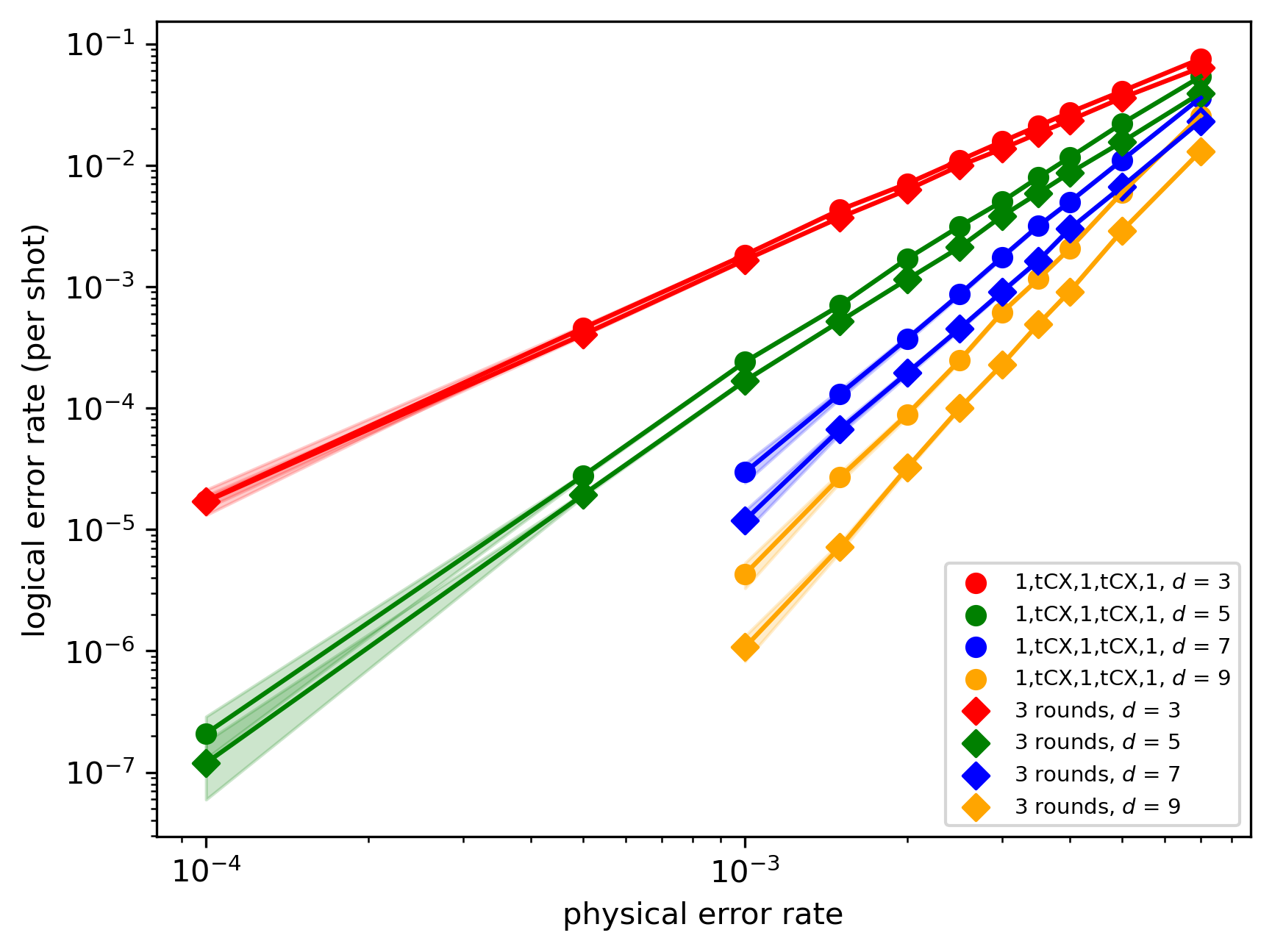}
    \caption{\label{fig:1_tCX01_1_tCX10_1} Surface code simulations with two alternating transversal CNOTs, one round of syndrome extraction between transversal CNOTs.}
\end{figure}

In this section we investigate the performance of the iterative CNOT decoder as a function of the number of rounds of syndrome extraction between successive alternating transversal CNOT operations. We show that in figure \ref{fig:1_tCX01_1_tCX10_1} where only a single round of error correction is performed between, we still maintain the same scaling relationship with physical error rate, and a comparable threshold. We note that the discrepancy between the transversal CNOT case versus memory appears to grow with increasing code distance. 

\begin{figure}[!h]  
    \centering
    \includegraphics[width=1\linewidth]{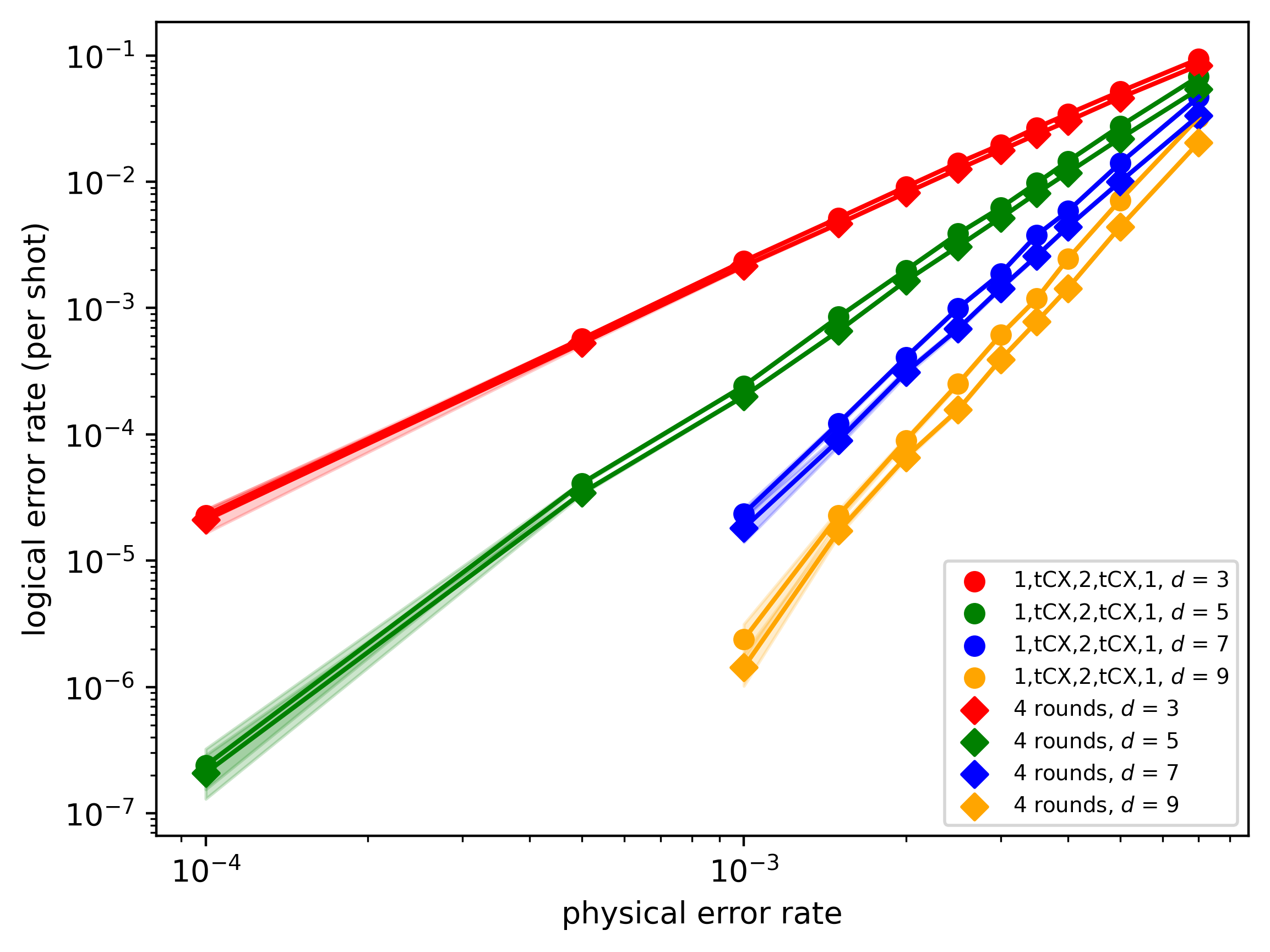}
    \caption{\label{fig:1_tCX01_2_tCX10_1} Surface code simulations with two alternating transversal CNOTs, two rounds of syndrome extraction between transversal CNOTs.}
\end{figure}

For figure \ref{fig:1_tCX01_2_tCX10_1} we instead include two rounds of syndrome extraction, and now this growing discrepancy with increasing distance is no longer evident. This difference may be attributable to the space-time edges that span across the syndrome extraction rounds where there is a degenerate choice of round occurrence in relationship to the location of the CNOT operations (discovered independently by \cite{SahayYT2024,sahay2024errorcorrectiontransversalcontrollednot}). The impact of this is mitigated by moving to two rounds and beyond. 

\begin{figure}[!h]  
    \centering
    \includegraphics[width=1\linewidth]{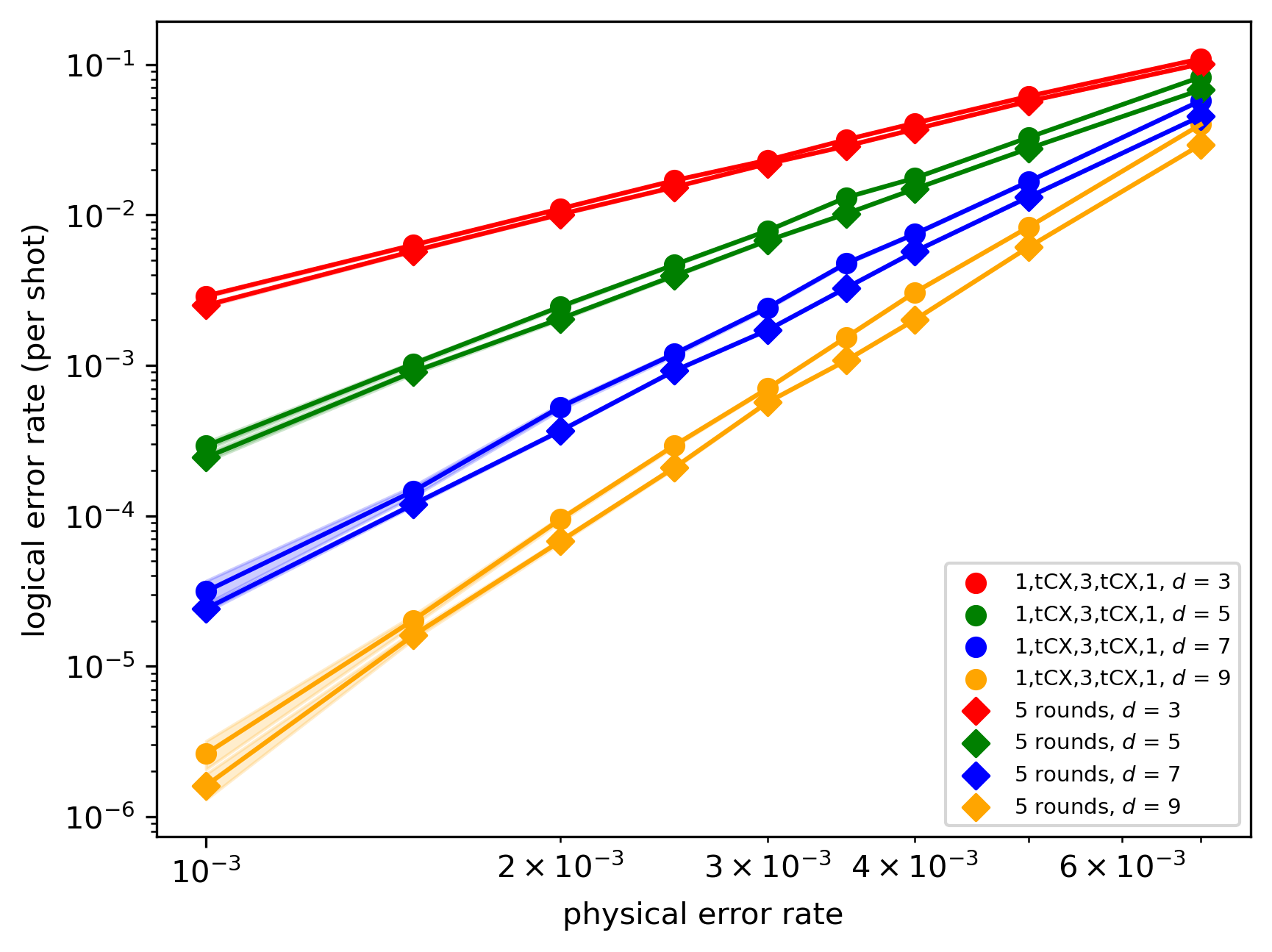}
    \caption{\label{fig:1_tCX01_3_tCX10_1_no_1e_minus_4}  Surface code simulations with two alternating transversal CNOTs, three rounds of syndrome extraction between transversal CNOTs.}
\end{figure}

In figure \ref{fig:1_tCX01_3_tCX10_1_no_1e_minus_4} we plot for the three round case, and see the same overall scaling relationships as the two round case. Of course, the ideal performance of a transversal CNOT decoder would be the minimal time complexity with the minimal final error rate. Moving from one round to two rounds improved our final logical error rate, e.g. for distance $9$, physical error rate $1$E-$3$, the final error rate improved from $4.3$E$-6$ to $2.4$E$-6$. In contrast, moving from two rounds to three rounds actually slightly worsened our final error rate, i.e. from $2.4$E$-6$ to $2.6$E$-6$. This is because there are opposing forces, i.e. the iterative propagating decoder has some additional resiliency to time component edges with increasing round count separating CNOT operations, but each additional round added itself has some additional contribution of error. We find a weak distance dependent effect for the optimal separating round count between alternating CNOT operations. This appears to be a unique property of alternating CNOT operations between two logical qubits and the optimal round count in this case will be a function of both the underlying physical error rate and the code distance, and for our investigations so far this is always substantially lower than the code distance, e.g. $2$ rounds for distance $9$ at an error rate $1$E-$3$. In the following section, we investigate the convergence of logical error rate with iteration number, and also include an example of a large relevant circuit where $1$ QEC round between layers of CNOT operations is shown to be optimal (i.e. matching memory equivalence). The optimal choice will ultimately include the desired target error rate, and a minimization of total volume, i.e. the distance required, and the number of rounds between CNOT operations to reach the target.

\subsection{Decoder convergence and more complex logical circuits}
Figure \ref{fig:y_fac} shows the pure-CNOT circuit component of a $\ket{Y}$ state distillation factory \cite{wan2024constanttimemagicstatedistillation,Herr_2017,zhou2024algorithmicfaulttolerancefast} with the state initialisation, $\ket{Y}$ state consumption and post-selection omitted. We simulated the logical error rate of this logical transversal CNOT circuit by computing its stabiliser flows \cite{Bombin_2024,McEwen_2023}.

\begin{figure}[!h]
    \centering
    \subfloat[][\label{fig:2_a_cx}A logical circuit of $2$ alternating transversal CNOTs.]{
    \resizebox{0.6\linewidth}{!}{

    }
    }
    \\
    \bigskip
    \subfloat[][\label{fig:3_a_cx}A logical circuit of $3$ alternating transversal CNOTs.]{
    \resizebox{0.8\linewidth}{!}{
    %
    }
    }
    \\
    \bigskip
    \subfloat[][\label{fig:4_a_cx}A logical circuit of $4$ alternating transversal CNOTs.]{
    \resizebox{0.9\linewidth}{!}{
    %
    }
    }
    \caption{The logical circuits simulated in figure \ref{fig:CONVERGENCE_PLOT}. The number/variable above every blue dotted line is the number of SE rounds between CNOTs.}
\label{fig:alterning_CNOTs}
\end{figure}

\begin{figure}[!h]
    \centering
    \resizebox{0.9\linewidth}{!}{
    %
    }
    \caption{The pure CNOT portion of the 7-to-1 $\ket{Y}$ state distillation circuit from \cite{wan2024constanttimemagicstatedistillation}, without the $\ket{Y}$ state consumptions. The number/variable above every blue dotted line is the number of SE rounds between CNOTs.}
\label{fig:y_fac}
\end{figure}

\begin{figure}[!h]
    \centering
    \resizebox{0.99\linewidth}{!}
    {\includegraphics[]{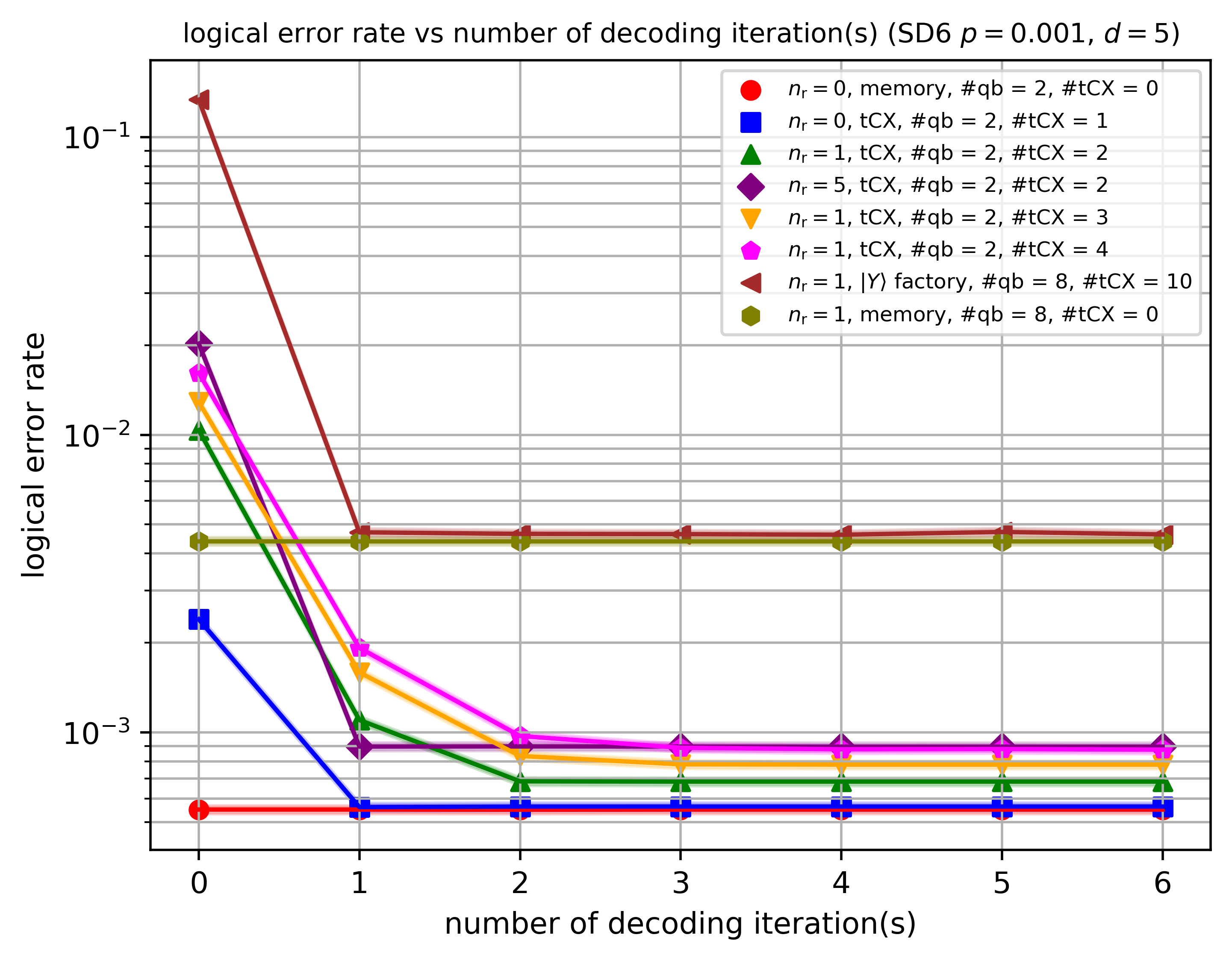}}
    \caption{Surface code simulations investigating the required number of iterations of the decoding method to reach optimal logical circuit quality for various circuits. Chain lengths of increasing CNOT count in an alternating arrangement on two logical qubits are investigated, starting from $0$ and ending in a length of $4$ CNOTs. The $\ket{Y}$ state distillation CNOT circuit is compared to a memory equivalent circuit. The number of QEC rounds between layers of CNOT gates here is stated for each trend.}
\label{fig:CONVERGENCE_PLOT}
\end{figure}

Figure \ref{fig:CONVERGENCE_PLOT} illustrates how the number of decoding iterations grows when we increase the chain length of consecutive, \textit{alternating} transversal CNOTs. We see that increasing the chain length of the alternating CNOT operations, requires a larger number of iterations to reach logical error rate convergence. In contrast, if a sufficient number of code cycles are introduced between successive alternating CNOTs, a single decoding iteration suffices to converge. This iteration dependence is unique to local alternating CNOT operations.

Moreover, for those \textit{alternating} circuits in Figure \ref{fig:CONVERGENCE_PLOT} where the logical error rate grows with CNOT number, this is directly related to each CNOT being followed by exactly one code cycle. In other words, we have added one full syndrome‐extraction cycle per CNOT operation. Nonetheless, if we compare \textit{each} of these curves to a corresponding memory‐equivalent baseline (not shown for every circuit), we find that after \textit{sufficient decoding passes} the logical error rate \textit{does} match that of pure memory - minimising any logical error rate overhead from propagated errors.

To emphasize the practical impact, we also studied a higher depth and logical qubits count, physically relevant $\ket{Y}$ state distillation subroutine \cite{Herr_2017,wan2024constanttimemagicstatedistillation,zhou2024algorithmicfaulttolerancefast}. Although it consists of ten transversal CNOT gates, none of them are local-and-alternating between any two logical qubits, and so one decoding iteration – and a single mid-circuit round between layers – was all that was required. Notably, the resulting logical error rate is (nearly) \textit{indistinguishable} from a pure memory experiment. Because we insert only one mid‐circuit round between layers of CNOT gates, the code-cycle-time overhead \textit{per gate} is then below $1$. In particular, the two total syndrome‐extraction rounds across the entire block of ten gates yields only $0.2$ code cycles per individual CNOT. Taken together, these results confirm that our iterative decoder imposes minimal overhead both in qubit-time complexity and decoding-time complexity for large relevant CNOT circuits. When the resulting logical error rate matches that of the equivalent memory experiment, it indicates that (nearly) \textit{all} extra errors from CNOT‐mediated propagation have been mitigated.

Our circuits are motivated from standard Stim library surface codes, with additional logic for circuit generation to handle multiple logical qubits and transversal CNOT operations. We do not include any unique detectors in our circuit beyond the standard memory equivalent circuits, i.e. there are no detectors conditioned on multiple logical qubits. This highlights a feature of our method, in that each logical qubit is decoded with MWPM independently, with only our additional logic on top to handle the propagation of errors. While the numerical results we present rely on decoding the syndrome graph in full, we believe our method is amenable to a windowed streaming decoder which we present in appendix \ref{appendix:streaming}, but we leave the implementation for future work.

\section{Discussions and outlook}
Our iterative decoder addresses the challenge of managing correlated errors resulting from transversal CNOT operations. The relevance of this work is supported by the feasibility of long-range two-qubit gates between physical qubits, as demonstrated in hardware platforms such as trapped ions \cite{webber2020efficientqubitroutingglobally,ryananderson2022implementingfaulttolerantentanglinggates,Bluvstein2024}.

A key outcome of this study is the finding that our decoding technique, which relies on MWPM, is capable of processing transversal CNOT circuits with sufficient quality to match memory-equivalent circuits. We observe a weak distance-dependent effect on the optimal number of QEC rounds separating alternating CNOT operations between two logical qubits; for example, $2$ QEC rounds were optimal for distance $9$, whereas $1$ round of QEC was optimal for distance $3$. This contrasts with a relevant large-scale example circuit ($\ket{Y}$ state distillation), where we found that just $1$ QEC round separating layers of CNOT gates matches the memory-equivalent circuit, resulting in a code cycle overhead of only $0.2$ per CNOT in this case. These results support the claim that transversal CNOT circuits can be decoded adequately with just $\mathcal{O}(1)$ QEC rounds per CNOT. We investigated the decoding complexity of our method—specifically, how logical error rate converges with iteration number (each loop through the CNOT circuit)—and again found unique behaviour for local alternating CNOT operations, where increasing the chain length raises the number of iterations required to optimize fidelity. In contrast, the $\ket{Y}$ state distillation circuit converged to match memory performance after only $1$ iteration. These results affirm the feasibility of our method from a convergence perspective, demonstrating its effectiveness even with a limited ($1$) iteration number in a relevant large circuit. Our results show that the iterative decoder maintains the expected scaling of logical error rates with code distance and physical error rates under typical circuit-level noise models.

A thorough investigation into the error mechanisms of local (i.e. not separated by sufficient code cycles, some value less than $d$) alternating CNOT circuits represents an interesting future research direction. In addition, a sketch on how to generalise the iterative transversal CNOT decoder to allowing for streaming syndrome data is available in appendix \ref{appendix:streaming}.

The practical implications of these findings are significant for particular quantum computing platforms. Incorporating transversal CNOT gates and the specialized decoding technique can reduce the physical qubit overhead required for fault-tolerant operations. For example, the lattice surgery CNOT implementation has a volume complexity of $\mathcal{O}(d^3)$, whereas the transversal CNOT has a volume complexity of $\mathcal{O}(d^2)$. The integration of transversal CNOTs would enhance the efficiency of magic state distillation processes and parallelized magic state consumption, thus reducing the overall resources required for quantum advantage applications in the fault-tolerant regime. This has the potential to bridge the gap in performance and resource requirements between different hardware architectures, such as superconducting qubits and trapped ions, making it feasible for slower, more connected systems to achieve comparable operational efficiency to their faster, locally-connected counterparts.

Detailed analysis of applying the iterative transversal CNOT decoder to more complicated logical circuits such as $\ket{Y}$ and magic $\ket{T}$ state distillation factories with post-selection is available at \cite{wan2024constanttimemagicstatedistillation}; showing promising logical performance and verifies the $\ket{Y}$ state distillation numerical results (decoded with a different decoder) from \cite{zhou2024algorithmicfaulttolerancefast}. Further exploration of hardware-specific error models and their impact on the performance of the transversal CNOT is essential. A detailed analysis of the performance of the transversal CNOT with hardware-realistic connectivity error models, and its contrast to the lattice surgery alternative, is in preparation \cite{tvslscnot}. Additionally, an extension to the quantum advantage resource estimation process to include the benefits of transversal CNOT operations is also in preparation \cite{tvslscnot}. In this upcoming work, physical qubit requirements to reach target runtime will be estimated for different connectivity and code cycle assumptions.

In summary, the multi-pass iterative decoder for transversal CNOT gates offers an approach to utilize the benefits associated with high-fidelity long-range qubit connectivity. By optimizing the number of QEC rounds required and the resulting error rate, this method can improve the performance of quantum hardware with such connectivity available.

\section{Acknowledgments}
We want to thank Niel De Beaudrap for his insights into analysing the iterative nature of the transversal CNOT decoder. We also acknowledge discussions with Craig Gidney, Michael Newman and Lucy J. Robson. We also want to thank all our colleagues at Universal Quantum Ltd for their continued support. Part of the numerical results were presented at the ICAP 2024 Satellite Conference (Brighton, UK) on 22 July 2024. Winfried K. Hensinger acknowledges support from Innovate UK (project number 10004857) and the University of Sussex. The idea of modifying the iterative CNOT decoder to allow for streaming syndrome data (appendix \ref{appendix:streaming}) originates from Austin Fowler.

\bibliography{main}
\clearpage
\onecolumn
\appendix

\section{Surface code implementation and Pauli frame inference}
\label{appendix:surface_code_implementation_and_pauli_frame_ingerence}
\subsection{Surface code implementations}
\begin{figure}[!h]
    \centering
    \begin{tikzpicture}
    \begin{scope}[xshift=0]
        \begin{yquant}
            qubit {} q[+1];
            qubit {} q[+1];
            qubit {} q[+1];
            qubit {} q[+1];
            qubit {$\ket{0}$} q[+1];
        
            H q[4];
            cnot q[0] | q[4];
            cnot q[1] | q[4];
            cnot q[2] | q[4];
            cnot q[3] | q[4];
        
            H q[4];
        
            measure q[4];
        \end{yquant}
    \end{scope}
    \begin{scope}[xshift=200]
        \begin{yquant}
            qubit {} q[+1];
            qubit {} q[+1];
            qubit {} q[+1];
            qubit {} q[+1];
            qubit {$\ket{0}$} q[+1];
        
            cnot q[4] | q[0];
            cnot q[4] | q[1];
            cnot q[4] | q[2];
            cnot q[4] | q[3];
                
            measure q[4];
        \end{yquant}
    \end{scope}
    \end{tikzpicture}
    \caption{\label{fig:SE_circuits} X (left) and Z (right) stabilizer measurement circuits for the surface code.}
\end{figure}
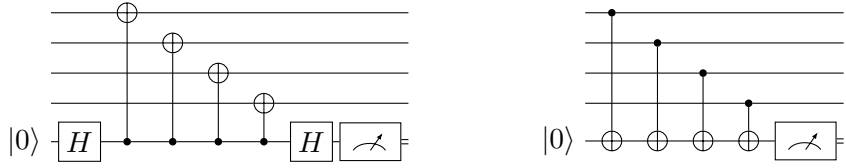

\begin{figure}[!h]
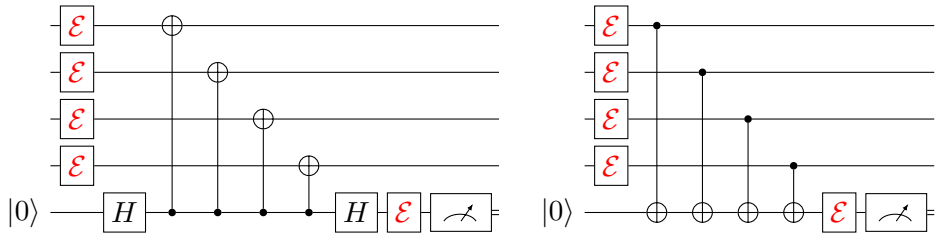

    \centering
    %
    \caption{\label{fig:SE_circuits_pheno_err_model}Syndrome extraction circuits for the surface code with a phenomenological error model, a stochastic Pauli error ({\color{red}$\mathcal{E}$}) can occur only before the syndrome extraction circuit and before the measurement and not mid-circuit.}
\end{figure}

We implement our surface code simulations in the following manner, we perform our yellow (X) and grey (Z) parity checks, with the left and right syndrome extraction circuits respectively in figure \ref{fig:SE_circuits}, in the order dictated by the black and white arrows as show in figure \ref{fig:SE_CX_ordering}.
We can draw the Tanner graph of a single time slice as shown in figure \ref{fig:blue_tanner} for the Z parity checks (detecting X errors) Tanner graph lattice, where circular nodes represents qubits and square nodes indicating detectors. The matching graph is equivalent to this Tanner graph with all the circular nodes removed and represented as edges. We can join multiple of these times slice together in the $z$ dimension, linking square detector nodes to create the standard matching graph used to decode for the phenomenological error model (see figure \ref{fig:SE_circuits_pheno_err_model}). In this case vertical edges now indicate measurement errors in the syndrome extraction circuit.

\begin{figure}[!h]
     \centering
     \subfloat[][Ordering of physical CNOT gates performed to carry out syndrome extraction \cite{PhysRevA.90.062320,Yoder_2017}.]{
     \resizebox{0.28\linewidth}{!}{

} \label{fig:SE_CX_ordering}}
     \hspace{0.1cm}
     \subfloat[][Single time slice of the Tanner graph with the grey (Z) parity checks represented as square nodes. Orange nodes indicate the same boundary node. A matching graph is the the same graph with all the circular qubit nodes dropped, and the edges remaining will represent a potential error.]{
     \resizebox{0.28\linewidth}{!}{
    %
    } \label{fig:blue_tanner}}
     \hspace{0.1cm}
     \subfloat[][A standard phenomenological error model matching graph for the rotated surface code (Z parity checks that detect X errors).]{
     \resizebox{0.28\linewidth}{!}{
    %
    } \label{fig:}}
     \caption{}
     \label{fig:}
\end{figure}

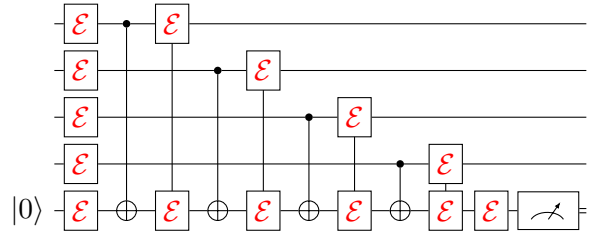
\begin{figure}[!h]
    \centering
    \begin{tikzpicture}
    \begin{scope}[xshift=0]
        \begin{yquant}
            qubit {} q[+1];
            qubit {} q[+1];
            qubit {} q[+1];
            qubit {} q[+1];
            qubit {$\ket{0}$} q[+1];

            [text = red] box {$\mathcal{E}$} q[0]-q[4];
            
            align q;
            cnot q[4] | q[0];
            [text = red, name = box0] box {$\mathcal{E}$} q[0],q[4];
            cnot q[4] | q[1];
            [text = red, name = box1] box {$\mathcal{E}$} q[1],q[4];
            cnot q[4] | q[2];
            [text = red, name = box2] box {$\mathcal{E}$} q[2],q[4];
            cnot q[4] | q[3];
            [text = red, name = box3] box {$\mathcal{E}$} q[3],q[4];

            [text = red] box {$\mathcal{E}$} q[4];
            measure q[4];          

        \end{yquant}
    \end{scope}
    \draw (box0-0) -- (box0-1) (box1-0) -- (box1-1)  (box2-0) -- (box2-1)  (box3-0) -- (box3-1);

    \end{tikzpicture}
    \caption{\label{fig:SE_circuits_cir_level_err_model} A syndrome extraction circuit for the surface code with a circuit-level noise, a stochastic Pauli error ({\color{red}$\mathcal{E}$}) can occur before the round, before the measurement and after every operation.}
\end{figure}

If we allow for circuit-level noise (see figure \ref{fig:SE_circuits_cir_level_err_model}), whereby every operation can be corrupted along with the operations outlined in the phenomenological error model, then the matching graph needs to have diagonal edges in spacetime to accommodate and to identify these mid-circuit errors. Some small number of these diagonal spacetime edges, along with some purely temporal edges are included in figure \ref{fig:mid_circuit_decompose} for illustrations purposes.

\subsection{Pauli frame inference}
\begin{figure}[!h]
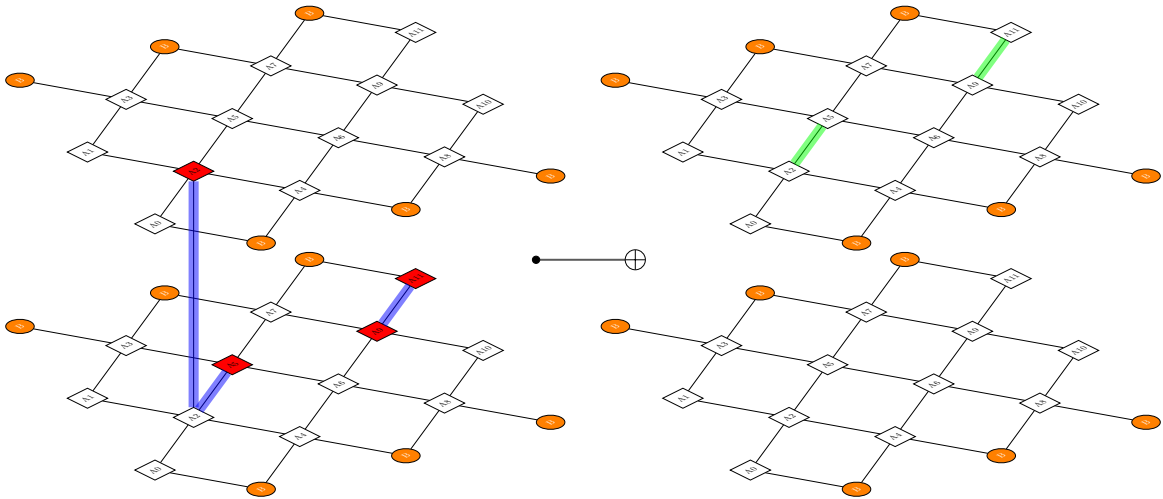

    \centering
    \resizebox{0.9\linewidth}{!}{

    } 
    \caption{\label{fig:pauli_frame_propagation} Propagation of error edges across a transversal CNOT. Note that only some spacetime and purely temporal edges are shown for a clearer representation. }
\end{figure}

Following similar notations and colour scheme as shown in the repetition code example in the main text, we shall illustrate how we propagate the Pauli frame in the surface code. We shall restrict to the Z parity check lattice 
 that detects X errors only, but this should work for the X parity checks as well by symmetry. We run MWPM on this matching graph to obtain blue shaded edges via matching. If a transversal CNOT is applied after the syndrome extraction circuits leading to the detectors in the bottom layers in figure \ref{fig:pauli_frame_propagation}, following similar arguments, we propagate the blue decoded edges from the left to the right logical qubit as green edges.

\begin{figure}[!h]
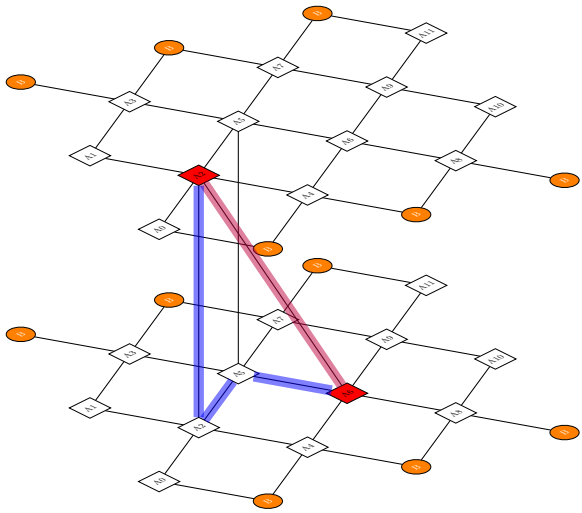
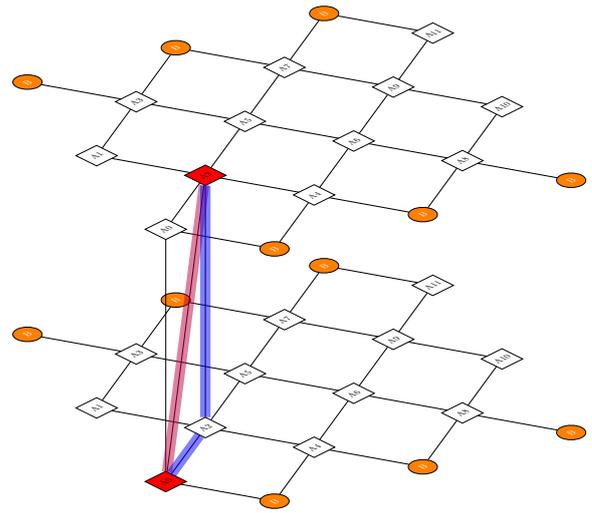

    \centering
    \subfloat[][\label{fig:decompose_to_weigh_3} Decomposition of a diagonal spacetime edge into a weight 3 error chain purely composed of spatial (weight 2) and temporal edges (weight 1).]{     
    \resizebox{0.45\linewidth}{!}{

    } }
    \hspace{1cm}
    \subfloat[][\label{fig:decompose_to_weigh_2} Decomposition of a diagonal spacetime edge into a weight 2 error chain purely composed of spatial (weight 1) and temporal edges (weight 1).]{
    \resizebox{0.45\linewidth}{!}{
    
    %
        } }
    \caption{\label{fig:mid_circuit_decompose} Decomposition of diagonal edges representing mid-circuit errors to purely spatial and temporal edges for Pauli frame propagation purposes. The left and right matching graphs are two such examples. Only some of the purely temporal and spacetime diagonal edges are shown for a clearer representation.}
\end{figure}

If mid-circuit error edges had been decoded and identified as the purple shaded edges in both sub-figures in figure \ref{fig:mid_circuit_decompose}, we take a higher weight error chain as shown by the highlighted blue coloured edges in figures \ref{fig:decompose_to_weigh_3} and \ref{fig:decompose_to_weigh_2}. We only propagate the Pauli frame as shown by the purely blue coloured spatial edges after decomposition and ignore the purely temporal edges.

\clearpage

\section{Pseudo-code for $m$ alternating CNOTs between two patches}
\twocolumn
\label{appendix:psuedocode_2_patch}
\begin{figure}[!h]
\centering
\resizebox{0.95\linewidth}{!}{
    %
    }
\caption*{}
\end{figure}

In this section, we show the pseudo-code for the algorithm used to generate the numerical results in the main text, here are the following notations:
\begin{itemize}
    \item Binary matrix $S_{i,\alpha}^{k,l}$ representing the Z-check syndromes at the $i^{\text{th}}$ syndrome extraction, detector index $\alpha$, for logical qubit $k$, after $l$ iterations. For the X-checks, reverse target and control.

    \item Binary matrix $B_{i,j}^{k,l}$ representing the presence of X-errors right before the $i^{\text{th}}$ syndrome extraction, physical qubit index $j$, for logical qubit $k$, after $l$ iterations.

    \item Binary tensor $G_{i,j}^{k,l}$ representing the presence of X-errors right before the $i^{\text{th}}$ syndrome extraction, physical qubit index $j$, for logical qubit $k$, after $l$ iterations.

    \item CNOT direction vector: $\vec{C} = [01,10,01,10,01,...,01]$ stores the information of which qubit the control and target is acting on. $01$ means the CNOT at that position is acting on logical qubit $0$ as control and logical qubit $1$ as target.

    \item Parity check round numbers vector: $\vec{x} = [x_0,x_1,...,x_{m-1},x_{m}]$, stores the number of parity check rounds between CNOTs.

    \item $l_{\text{max}}$ is the maximum number of iterations, a free parameter.
\end{itemize}

Not that $B_{i,j}^{k,l}$ is a matrix ranging over only the subscripts indices $(i,j)$ and labelled by $k,l$ as the $l^{\text{th}}$ iteration and $k^{\text{th}}$ logical qubit. Note that the $B_{i,j}$ object is a matrix of many numbers. Also, we slice the matrix in this manner: $B_{i=0:7,j}^{k,l}$, indicating the first 8 elements of $i$ is sliced. Summation in modulo 2 is represented as $\bigoplus_{i=0}^{7} B_{i,j}^{k,l}$, summing the first 8 rows in modulo 2. 

\begin{algorithm}[H]
    \scriptsize
    \label{algo:2_tCX_alt}
    
    \begin{algorithmic}[1]
      
        \Input{Syndromes $S_{i,\alpha}^{k,0}$, zero matrices $B_{i,j}^{k,0},G_{i,j}^{k,0}$, $\vec{C}$, $\vec{x}$, $l_{\text{max}}$ }
        \Output{$B_{i,j}^{k,l},G_{i,j}^{k,l}$}
        
        \Function{MWPMDecode}{$S_{i,\alpha}^{k,l}$}
            \State Decode $S_{i,\alpha}^{k,l}$ via MWPM.
            \State \Return $D_{i,j}^{k,l}$ (decoded values)
        \EndFunction

        \Function{ErrorToDet}{$D^{k,l}_{i,j}$}
            \State Generate detector pattern associated with errors: $D^{k,l}_{i,j}$.
            \State \Return $S'$
        \EndFunction

        \Function{IterativeDecode}{$S_{i,\alpha}^{k,0}$}
            \State Termination Condition $\gets$ False, $l \gets 0$
            \State $\mathcal{G}^{k}_{z,j} \gets 0 \ \forall z,j,k$ (remove previous propagation)
            \State $\mathcal{S}^{k}_{z,j} \gets 0 \ \forall z,j,k$ (remove previous propagation)

            \State DecodeNext $\gets$ True
            
            \While {Termination Condition = False}
                \For {$a = 0$ \To $m - 1$} (for every CNOT)
                    \State Compute $y = \sum_{\gamma = 0}^{a} x_{\gamma}$ (SE before $a^{\text{th}}$ CNOT)
                    \If{$C_a=01$}
                        \State $c \gets 0, \ t \gets 1$
                    \Else
                        \State $c \gets 1, \ t \gets 0$
                    \EndIf

                     \If{DecodeNext = True}

                        \State $B_{i,j}^{k=c,l} \gets $ \Call{MWPMDecode}{$S_{i,\alpha}^{k=c,l}$}
                    \EndIf
                    
                    \State $\mathcal{P}_{\text{frame}} \gets \bigoplus_{i = 0}^{y-1}B_{i=0:(y-1),j}^{k=c,l}\oplus G_{i=0:(y-1),j}^{k=c,l}$
                    
                    \State $S' \gets$ \Call{ErrorToDet}{$\mathcal{P}_{\text{frame}}$} 

                    \If{$S' \neq $ zero tensor}
                        \State DecodeNext $\gets$ True
                    \Else{}
                        \State DecodeNext $\gets$ False
                    \EndIf
                    
                    \State $S_{i=y,\alpha}^{k=t,l} \gets S_{i=y,\alpha}^{k=t,l} \oplus S'\oplus\mathcal{S}^{k=t}_{z=a,j}$ (spurious events)

                    \State $\mathcal{S}^{k=t}_{z=a,j} \gets S'$ 

                    \State $G_{i=y,j}^{k=t,l} \gets G_{i=y,j}^{k=c,l} \oplus \bigoplus_{i = 0}^{y-1}B_{i=0:(y-1),j}^{k=c,l}\oplus \mathcal{G}^{k=t}_{z=a,j}$
                    
                    \State $\mathcal{G}^{k=t}_{z=a,j} \gets G_{i=y,j}^{k=t,l}$ 

                \EndFor
                \State $l \gets l+1$
                \State $B_{i,j}^{k,l},G_{i,j}^{k,l},S_{i=y,\alpha}^{k=t,l} \gets B_{i,j}^{k,l+1}, G_{i,j}^{k,l+1},S_{i=y,\alpha}^{k=t,l+1}$ 
                \If{$l = l_{\text{max}}$ or   $B_{i,j}^{k,l},G_{i,j}^{k,l} = B_{i,j}^{k,l-1},G_{i,j}^{k,l-1}$}
                
                    \State Termination Condition $\gets$ True
                \Else 
                    \State Termination Condition $\gets$ False 
                \EndIf 
            \EndWhile

            \State \Return $B_{i,j}^{k,l},G_{i,j}^{k,l}$
        \EndFunction

        \State $B_{i,j}^{k,l},G_{i,j}^{k,l} \gets$ \Call{IterativeDecode}{$S_{i,\alpha}^{k,0}$}
    
        \end{algorithmic}   
    \caption{\scriptsize $m$ alternating CNOTs between two patches (for X-errors)}
\end{algorithm}

\clearpage
\onecolumn
\section{Modifications to a streaming decoder}
\twocolumn
\label{appendix:streaming}
We illustrate a method to extend and modify the iterative transversal CNOT decoder to allow for a streaming syndrome data decoder. Figures \ref{fig:steaming_big_start} to \ref{fig:steaming_big_5} shows two logical qubits (left and right) with an infinite train of alternating logical CNOTs between them. This toy example can illustrates the worst case example of Pauli frame propagation back and forth between two logical qubits. 

We follow the same diagram conventions from the main text except two minor deviation from the notation in the main text:
\begin{itemize}
    \item vertex nodes from the main text are omitted and it's assumed that the vertices in the following matching graphs are detectors, 
    \item toggled-off DE are shown by a white node: \resizebox{0.2\linewidth}{!}{
    \begin{tikzpicture}[scale=1,every node/.style={minimum size=1cm},on grid]        
            \begin{scope}[xshift=0,every node/.append style={yslant=0,xslant=0},yslant=0,xslant=0]
            \fill[white,fill opacity=0.0] (0,0) rectangle (5,1);
            \node[fill=none,shape=circle,draw=none] (N0T0) at (0,0) {};
            \node[fill=white,shape=circle,draw=black] (N1T0) at (2,0) {};
            \node[fill=none,shape=circle,draw=none] (N2T0) at (4,0) {};
    
            \path [-,line width=0.1cm,black,opacity=1] (N0T0) edge node {} (N1T0);
            \path [-,line width=0.1cm,black,opacity=1] (N1T0) edge node {} (N2T0);
            
    \end{scope}
    \end{tikzpicture}
    }, and
    \item additional gray coloured shaded exploratory region around
the chosen vertex is inline with conventions from \cite{Fowler_2012_timing} (only available in their arXiv version 2, not in the PhysRevLett.108.180501 version).
\end{itemize}
Follow figures \ref{fig:steaming_big_start} to \ref{fig:steaming_big_5} in chronological order along with the figure sub-captions for a sketch modification of the iterative transversal CNOT decoder to accommodate for streaming syndrome data. A finite window \cite{acharya2024quantumerrorcorrectionsurface} can be set on how far to look for updates in the past. Numerical performance and verifications is left for future work.
\begin{figure}[!h]
    \centering
    \subfloat[\label{fig:streaming1} Streaming example, initially no errors or detection events (DEs). All the streaming syndrome data have been received from $t=0$ to $t=2$.]{{
    \resizebox{0.95\linewidth}{!}{

    }}}
    \\
    \bigskip
    \subfloat[\label{fig:streaming2} One error, one DE, match. We have yet to receive the streaming syndrome data from $t=4$.]{{
    \resizebox{0.95\linewidth}{!}{
    %
    }}}
    \caption{}
    \label{fig:steaming_big_start}
\end{figure}

\begin{figure}[!h]
    \centering
    \subfloat[\label{fig:streaming3} The syndrome data received up to $t=4$. Two DEs added (one from new error, one from propagation of unknown errors, layer by layer using the all history mod 2 of red and brown lines), propagate corrections (layer by layer using the all history mod 2 of blue and green lines), one DE toggled off (white).]{{
    \resizebox{0.95\linewidth}{!}{
    %
    }}}  
    \\
    \bigskip
    \subfloat[\label{fig:streaming4} Match $\Rightarrow$ new {\color{blue}blue} edges on the left logical qubit. Streaming syndrome data received up till $t=4$.]{{
    \resizebox{0.95\linewidth}{!}{
    %
    }}}  
    \\
    \bigskip
    \subfloat[\label{fig:streaming5} Propagate corrections, one DE toggled on. Streaming syndrome data received up till $t=4$.]{{
    \resizebox{0.95\linewidth}{!}{
    %
    }}}  
    \\
    \bigskip 
    \subfloat[\label{fig:streaming6} Match. Streaming syndrome data received up till $t=4$.]{{
    \resizebox{0.95\linewidth}{!}{
    %
    }}}  
    \caption{}
    \label{fig:steaming_big_1}
\end{figure}

\begin{figure}[!h]
    \centering   
    \subfloat[\label{fig:streaming7} Streaming syndrome data received up till $t=5$ now. Two DEs added (one from time-like error, one from propagation), propagate corrections, one DE toggled off (white).]{{
    \resizebox{0.95\linewidth}{!}{
    %
    }}}  
    \\
    \bigskip
    \subfloat[\label{fig:streaming9}Propagate corrections layer by layer undoing matching inconsistent with toggled DEs.]{{
    \resizebox{0.95\linewidth}{!}{
    %
    }}}  
    \\
    \bigskip
    \subfloat[\label{fig:streaming10}Advance time to $t=6$, new error, propagate errors (nothing to propagate as left qubit red and brown lines cancel and left blue and green lines cancel).]{{
    \resizebox{0.95\linewidth}{!}{
    %
    }}}  
    \caption{}
    \label{fig:steaming_big_2}
\end{figure}

\begin{figure}[!h]
    \centering
    \subfloat[\label{fig:streaming11}Match.]{{
    \resizebox{0.95\linewidth}{!}{
    %
    }}}  
    \\
    \bigskip
    \subfloat[\label{fig:streaming12}Propagate corrections layer by layer undoing matching inconsistent with toggled DEs.]{{
    \resizebox{0.95\linewidth}{!}{
    %
    }}}  
    \\
    \bigskip
    \subfloat[\label{fig:streaming14}Propagate corrections layer by layer undoing matching inconsistent with toggled DEs. Note currently unable to match one DE.]{{
    \resizebox{0.95\linewidth}{!}{
    %
    }}}  
    \caption{}
    \label{fig:steaming_big_3}
\end{figure}

\begin{figure}[!h]
    \centering
    \subfloat[\label{fig:streaming15}Advance time to $t=7$, propagate errors and corrections. Toggle DEs. Match.]{{
    \resizebox{0.95\linewidth}{!}{
    %
    }}}  
    \\
    \bigskip
    \subfloat[\label{fig:streaming16}Propagate, toggle DEs undoing matchings if necessary.]{{
    \resizebox{0.95\linewidth}{!}{
    %
    }}} 
    \caption{}
    \label{fig:steaming_big_4}
\end{figure}

\begin{figure}[!h]
    \centering
    \subfloat[\label{fig:streaming18}Advance time to $t=8$. Propagate, toggle DEs undoing matchings if necessary.]{{
    \resizebox{0.95\linewidth}{!}{
    %
    }}} 
    \caption{}
\label{fig:steaming_big_5}

\end{figure}

\end{document}